\documentclass[aps,prd,preprint,groupedaddress]{revtex4}

\usepackage{amsmath}
\usepackage{amsfonts}
\usepackage{graphicx}

\newcommand{\wdark}{w_{\rm DE}}
\newcommand{\wtot}{w_{\rm tot}}

\begin{document}


\title{Exploring extra dimensions through observational 
tests 
of dark energy and  varying Newton's constant}

\author{Paul J. Steinhardt$^{1,2}$ and Daniel Wesley$^{3,4}$}
\email[]{steinh@princeton.edu}
\email[]{dwes@sas.upenn.edu}
\affiliation{$^1$Joseph Henry Laboratories, 
Princeton University, Princeton, NJ 08544 \\
$^2$Princeton Center for Theoretical Science, 
Princeton University, Princeton, NJ 08544 \\
$^3$Centre for Theoretical Cosmology, DAMTP, Cambridge University, Cambridge, CB3 0WA, United Kingdom \\
$^4$Center for Particle Cosmology, 
Department of Physics, University of Pennsylvania, Philadelphia PA 19104-6395 USA}

\date{\today}

\begin{abstract}
    \noindent
   We recently presented a series of dark energy theorems that place 
    constraints on the equation of state of dark energy ($\wdark$),  the
    time-variation of Newton's constant ($\dot G$), and the violation of energy 
    conditions in theories with extra dimensions.  In this paper, we explore 
    how current and future measurements of $\wdark$ and $\dot G$ can be used to place tight limits on
     large classes of these theories (including some of the 
    most well-motivated examples) independent of the size of the extra dimensions.   As an example, we show that models with 
    conformally Ricci-flat metrics obeying  the null energy condition (a common 
    ansatz for 
    Kaluza-Klein and string constructions) are highly constrained by current 
    data and may be ruled out entirely 
    by future dark energy and pulsar observations. 
\end{abstract}


\maketitle

\section{Introduction}

       Beginning with the work of Kaluza and Klein \cite{KK1,KK2,KK3} and continuing today with string theory and M-theory, extra dimensions have been a common feature of unified theories.  The basic notion is that the observed 3+1-dimensional universe is actually described by a general relativistic theory in a space-time with one or more extra compactified dimensions.  If the compactification scale is much greater than 1~TeV (or $<<10^{-16}$~cm), laboratory experiments, even at the Large Hadron Collider, are unable to
       uncover direct evidence 
       of extra dimensions.
       
       In this paper, though, we show how measurements of the equation of state of dark energy ($\wdark$) and the time variation of Newton's constant $\dot G/G$ can be used to test or rule out the existence of extra dimensions for large classes of models {\it independent of the compactification scale.}  This surprising power to discriminate among extra-dimensional models derives from a set of ``dark energy theorems,"  first described in \cite{Wesley:2008de,Wesley:2008fg, Steinhardt:2008nk}.

         The theorems are based on the observation that the expansion of the usual three large dimensions tends to cause extra dimensions to vary with time, which, in turn, causes a change of $\wdark$ and $G$ in the corresponding 4d effective theory.  These changes can be avoided in a decelerating universe by introducing conventional interactions strong enough to keep the sizes of the extra dimensions fixed. However, the dark energy theorems show that, once the universe starts to accelerate, conventional interactions satisfying the classical (strong, weak and null) energy conditions no longer suffice no matter the size of the extra dimensions. Many well-motivated extra-dimensional models satisfy one or more of the energy conditions.  For these large classes, the dark energy theorems, combined with the observed acceleration rate, can be used to compute the predicted time-variation of $\wdark$ and $G$ for given model parameters.  As illustrated below, current measurements are already strong enough to rule out a substantial range of model parameters.  The more exciting prospect is anticipated improvements in the measurement of $\wdark$, as described by the Dark Energy Task Force (DETF), and of $\dot G/G$, as constrained by pulsar timing, that can test or rule out whole classes of extra-dimensional models.

       The dark energy theorems were derived for the general case of  $k$ extra spatial dimensions, but the predictions depend on $k$.    For the purposes of illustration, we focus in this paper on the well-motivated class of 
    9+1-dimensional theories ($k=6$) with a conformally flat Ricci (CRF) metric and satisfying the null energy condition (NEC) -- theories  commonly used in string- and M-theoretic phenomenological and cosmological  models. In Ref~\cite{Steinhardt:2008nk}, we showed that this class of models is inconsistent with the standard $\Lambda$CDM model which has $\wdark=-1$.    In fact, the dark energy theorems show that, the closer $\wdark$ is to $-1$, the more 
    rapidly
    the extra-dimensional volume and, hence, $G$ 
    must vary. Furthermore, 
    if the theory contains no mechanism for violating the
    null energy condition,
     $\wdark$ cannot remain close to $-1$ 
     for an extended period.   
    This means that improved 
    limits on $\dot G$, combined with
    ever-tightening bounds on the time-variation of 
    $\wdark$ from future experiments,  
    can progressively constrain or rule out
    this entire class 
    of extra-dimensional models.

A corollary of this analysis is that a dark energy mission,
even if it fails to find any time-variation of $\wdark$ and is consistent with $\wdark=-1$, 
can still be  highly informative because it 
would eliminate well-motivated  extra-dimensional models.
A second corollary is that a coordinated effort is needed.  The ambitious improvements in the measurements of $\wdark$ alone, as projected by the DETF,  or of $\dot G/G$ alone, as estimated from planned pulsar timing surveys, are not sufficient.  Each constrains some range of parameter space but leaves some substantial untested range.  The two approaches are complementary, though: by pursuing both to some degree, entire classes of extra-dimensional models can be tested and ruled out.

This paper is organized as follows. 
In Section \ref{s:basiceqs}, we 
introduce 
 the conformally Ricci-flat (CRF) class of
 extra-dimensional  models  used
 to exemplify our approach and 
review the constraints  imposed by the dark energy 
theorems on the {\it total} equation of state $\wtot$ and Newton's constant $G$.  {\it Here and throughout this paper, the symbol $\wtot$ refers to  all contributions to the energy density of the universe (matter, radiation, etc.), not just dark energy}. The current value is $\wtot =-0.74$ based on observations \cite{Komatsu:2010fb}.  We use the symbol $\wdark$ to refer to the  dark energy component alone.

In Section \ref{s:method}, we 
show how to translate these constraints into 
predictions for dark energy and pulsar timing experiments.  The key results are in   
Section \ref{s:results}, where we  
compare the predictions with
current measurements and 
 near-future experiments.
In particular, we show how  current 
measurements of dark energy and pulsar timing
and anticipated improvements
can be used to test
and perhaps rule out the entire class of models.  
In
Section \ref{s:conclusions}, we conclude with a discussion of  the generalization to other classes of extra-dimensional models. More details on the constraints on the time-variation of $G$ and the dark energy equation of state in extra-dimensional models are given in the 
Appendices.


\section{\label{s:basiceqs} Dark Energy Theorems and Basic Equations}

In this section, we present a brief review of the ``dark energy theorems" 
first described and proven in \cite{Wesley:2008de,Wesley:2008fg, Steinhardt:2008nk}.
The theorems impose constraints on extra-dimensional theories in which the four-dimensional effective theory undergoes cosmic acceleration ($\wtot <-1/3$).
      The theorems assume a spacetime metric of the form
\begin{equation}\label{e:TDmetric-1}
\text{d} s^2 = g_{MN} \text{d} X^M \text{d} X^N = e^{2\Omega(t,y)} g^{\rm FRW}_{\mu\nu}(t,x) \, \text{d} x^\mu \text{d} x^\nu + h_{\alpha\beta}(t,y) \, \text{d} y^\alpha \text{d} y^\beta
\end{equation}
Here, $\mu,\nu$ ... are indices along the four large spacetime dimensions with coordinates $x^\mu$, and $\alpha,\beta$ ... indices along the $k$ compact dimensions with coordinates $y^\alpha$, and $N,M$ ... assume values along both compact and noncompact dimensions.  In (\ref{e:TDmetric-1}), $g^{\rm FRW}_{\mu\nu}$ is a flat Friedmann-Robertson-Walker metric, 
which we take to be
\begin{equation}
g^{\rm FRW}_{\mu\nu} = - {\cal N}(t)^2 \text{d}t^2
+ {\cal A}(t)^2 \delta_{mn} \text{d} x^m \text{d} x^n
\end{equation}
where $m,n$ range over $1 ... 3$.
The extra-dimensional metric $h_{\alpha\beta}$ and warp factor $\Omega$ can be time-dependent. 
For the purposes of illustration, we restrict ourselves in this paper to ``conformally Ricci-flat" (CRF) metrics:
 \begin{equation}
h_{\alpha \beta}(t,y) = e^{-2 \Omega (t,y)} h^{\rm RF}_{\alpha\beta} (t,y)
\end{equation}
where $h^{\rm RF}_{\alpha\beta} (t,y)$ has vanishing Ricci scalar, and $\Omega$ is the same warp factor which appears in (\ref{e:TDmetric-1}).  The CRF metric occurs in many string theory models such as warped Calabi-Yau \cite{Giddings:2001yu} and warped conifold \cite{Klebanov:2000hb} constructions
(where they are sometimes referred to as conformally Calabi-Yau metrics).  

In this work, we further assume that the extra-dimensional matter satisfies the null energy condition (NEC):
\begin{equation}\label{eq:NEC_def}
T_{MN} n^M n^N \ge 0
\end{equation}
for any null vector $n^M$, where $T_{MN}$ is the stress-energy tensor. 
For perfect fluids in four dimensions, the NEC 
requires that $\rho + P \ge 0$.   The NEC is satisfied by scalar 
fields with canonical kinetic terms (regardless of their potential), de 
Sitter and anti-de Sitter cosmological constants, $p$-form fields, and 
positive-tension extended objects.  To violate the NEC requires exotic 
ingredients such as scalar fields with higher-derivative kinetic energy terms or negative-tension 
extended objects (such as orientifold planes). Very often violation of the NEC 
leads to ghosts, instabilities, problems 
with gravitational thermodynamics, and other pathologies.

We will refer to the spectrum of extra-dimensional models having CRF metrics and obeying the NEC as the {\it NEC/CRF family of models}.  For concreteness, we will take the number of extra dimensions to be $k=6$.

As shown in \cite{Wesley:2008de,Wesley:2008fg, Steinhardt:2008nk}, 
constraints can be found by considering extra-dimensional theories for which the 4d effective theory is described by Einstein gravity and has $\wtot < -1/3$,  and, then,   following the overall expansion and dilation of the 
extra-dimensional metric in the extra-dimensional Einstein equations.  
The time evolution of the higher-dimensional metric can be expressed as the 
combination of a trace part $\xi$ and a symmetric, traceless shear component $\sigma_{\alpha \beta}$ \cite{Wesley:2008de,Wesley:2008fg, Steinhardt:2008nk}:
\begin{equation}
\frac{1}{2}
\frac{ \text{d} h_{\alpha \beta} }{\text{d} t} = 
\frac{1}{k} \xi h_{\alpha \beta}
+ \sigma_{\alpha \beta}
\end{equation}
where $h^{\alpha\beta} \sigma_{\alpha\beta} = 0$ and $h_{\alpha \beta}$,
$\sigma_{\alpha \beta}$ and $\xi$ are functions of $t$ and $y$.  The variable
$\xi$ is the local expansion rate of the extra-dimensional space.  
To measure the overall dilation of the extra dimensions, we define
a variable $\zeta_A$ by 
\begin{equation}\label{e:DefZeta}
\zeta_A = \frac{1}{H} \int e^{A\Omega} \xi \; \sqrt{h} \, \text{d}^k y
\end{equation}
where $H$ is the four-dimensional Hubble rate, and $A$ is a constant which
may be chosen for convenience.  Hence $\zeta_A$ represents the
fractional growth of the extra-dimensional volume per Hubble time, computed
using an $A$-dependent measure.  For the choice $A=2$, the volume measure in (\ref{e:DefZeta}) matches the one
which determines the four-dimensional Planck mass in warped compactifications.
Hence, for this value of $A$ we have
\begin{equation}\label{e:ZetaDotLnG}
\frac{\dot G}{G} = -H \zeta
\end{equation}
where $G$ is the four-dimensional Newton's constant, and 
we dropped the subscript $\zeta_2 \equiv \zeta$ for economy of notation.  When $\zeta$ is
nonzero, the volume of the extra dimensions is changing with time, and, 
hence, the four-dimensional Newton's constant is changing as well.

The dark energy theorems are derived by assuming that the
higher-dimensional matter fields satisfy the NEC \cite{Wesley:2008de,Wesley:2008fg, Steinhardt:2008nk}.  
By dividing the space-space components of the stress-energy tensor into two blocks corresponding the non-compact and compact
directions, two pressure-like parameters can be constructed by taking trace averages
over the $3\times 3$ and $k \times k$ blocks of the higher-dimensional
metric,
\begin{equation}
p_3 \equiv \frac{1}{3} g^{mn} T_{mn}
 \; \; {\rm and} \; \;
p_k \equiv \frac{1}{k} h^{\alpha\beta} T_{\alpha\beta},
\end{equation}
where $m,n$ range over the spatial coordinates $1...3$ and
$\alpha,\beta$ range over the extra-dimensional coordinates as in
(\ref{e:TDmetric-1}).
The NEC is violated if either $\rho+p_3$ or $\rho+p_k$ is less than zero at any space-time point, where $\rho \equiv -{T_0}^0$ is the higher dimensional energy density; or if the volume weighted average of either is less than zero; or if either of the ``A-weighted'' averages (the volume-weighted averages 
of 
$ e^{A \Omega}(\rho+p_3)$ or $e^{A \Omega}(\rho+p_k)$) is  less than zero for any $A$. By combining the higher-dimensional Einstein equations, expressions can be derived relating the $A$-weighted averages to $\xi$, $\sigma_{\alpha \beta}$, $\wtot$, $\Omega$, $k$ and the 4d effective energy density $\rho_{4d}$.  The condition that the $A=2$-weighted average of $\rho+p_k$ be non-negative can be rearranged into the constraint:    
\begin{equation}\label{e:ZetaDIE}
\frac{\text{d} \zeta}{\text{d} N} \ge \alpha_0 + \alpha_1 \zeta + \alpha_2 \zeta^2
\end{equation}
where $N = \ln (a)$ and $a$ is the Einstein frame scale factor.  The analogous condition for $\rho+ p_{3}$  is
\begin{equation}\label{e:ZetaID}
\zeta^2 \le F
\end{equation}
The functions $\alpha_0$, $\alpha_1$, $\alpha_2$, and $F$ depend on $\wtot$, the number of extra dimensions $k$ and the type of extra-dimensional metric.  
For our example with CRF metric
and  $k=6$ extra dimensions, the strongest observational constraints 
are found for $A=2$, in which case the
value of $\zeta$ is  related to the variation of Newton's constant in four dimensions by (\ref{e:ZetaDotLnG}).
Then, the constraint equations (\ref{e:ZetaDIE}) and (\ref{e:ZetaID}) required for the extra-dimensional theory to satisfy the Einstein equations and the NEC become \cite{Wesley:2008de,Wesley:2008fg, Steinhardt:2008nk}:
\begin{equation}\label{e:CRF6ZetaDIE}
\frac{\text{d} \zeta}{\text{d} N} \geq \zeta^2 + \frac{3(\wtot-1)}{2} \zeta - \frac{9(1+3 \wtot)}{4} 
\end{equation}
and
\begin{equation}\label{e:CRF6ZetaID}
\zeta^2 \leq \frac{9(1+\wtot)}{2}.
\end{equation}
With (\ref{e:CRF6ZetaDIE}) and (\ref{e:CRF6ZetaID}) in hand, we can derive some
qualitative consequences of the dark energy theorems for the NEC/CRF family of models. For example, suppose we wish to
find a solution  for which the four-dimensional Newton's constant does not vary ($\zeta = 0$). With this
choice of $\zeta$, the 
constraint equation 
(\ref{e:CRF6ZetaDIE}) can only be satisfied if $\wtot \ge -1/3$.  Conversely, if the four-dimensional universe is accelerating, then the $\zeta$ must be non-zero and the 4d Newton's constant must vary.

If $\wtot < -1/3$ and the universe is accelerating,  $\zeta$ cannot vanish but one might look for cases where $\wtot$ and $\zeta$ are constant.  Such solutions require that there exists
a value of $\zeta$ such that the right-hand side of (\ref{e:CRF6ZetaDIE})
vanishes.  This is only possible when $\wtot \geq -5 + 2\sqrt{5} \simeq -0.53$.
We will denote this value of $\wtot$ by $w_{\rm trans}$.  Thus, a second corollary of the dark energy theorems is that steadily accelerating
solutions with constant $\zeta$ are only possible when $\wtot \ge w_{\rm trans}$, which in this case corresponds to $\wtot \gtrsim -0.53$.

The most striking conclusions of all are reached by considering values of
$\wtot$ below $w_{\rm trans}$.   For $\wtot < w_{\rm trans}$, the right-hand side of
(\ref{e:CRF6ZetaDIE}) is positive definite for all values of $\zeta$, so
$\zeta$ must evolve with time. Since $\zeta$ is increasing with time,
 after a finite number of e-foldings the bound encapsulated in
 (\ref{e:CRF6ZetaDIE}) and (\ref{e:CRF6ZetaID}) will be violated.  Hence there are
\underline{no} steadily accelerating solutions for $\wtot < w_{\rm trans} \simeq -0.53$:
acceleration with $\wtot < w_{\rm trans}$ is necessarily transient.  


\section{\label{s:method} Observational constraints and the $w_0-w_a$ plane}

The two constraint equations Eq.~(\ref{e:CRF6ZetaDIE}) and~(\ref{e:CRF6ZetaID}) restrict the time-variation of $G$ and $\wtot$ for any extra-dimensional model described by a CRF metric and obeying the NEC.  In this section, we explain how to express these theoretical constraints as limits on the time-dependence of $\wdark$ and, then, how to incorporate observational constraints that can further restrict or perhaps rule out the remaining, theoretically allowed possibilities.

The first step is to parameterize the time-dependence 
of $\wdark$ and $\zeta$.  In general, each can have 
complicated time-dependence.  However, in keeping with standard practice, we 
take $\wdark$ to be a simple function of the scale factor $a$ over a range encompassing the present epoch ($0< a <2$ where $a\equiv 1$ today):  
\begin{equation}\label{e:DEQuad}
\wdark(a) = w_0 + (1-a) w_a + (1-a)^2 w_b.
\end{equation}
The quadratic form is used, rather than the linear DETF parameterization, because the analysis would produce artificially strong constraints for a pure linear dependence.

\begin{figure}
    \begin{center}
    \includegraphics[width=6in]{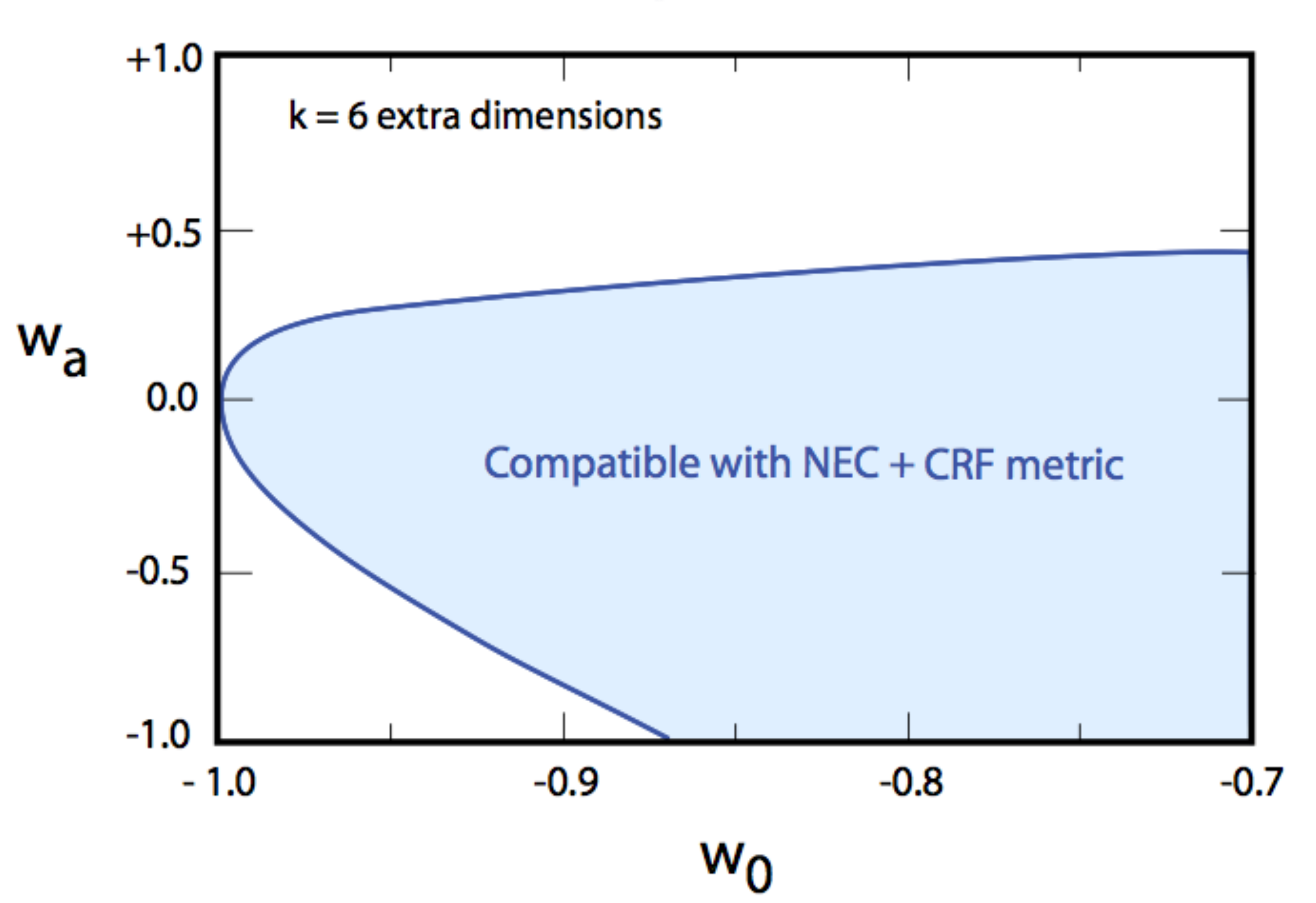}
    \caption{\label{f:NEC_v_SEC} The $(w_0,w_a)$-plane where $\wdark =w_0+ w_a (1-a) + w_b (1-a)^2$.  The plot shows the boundary of ${\cal C}$, the set of points   {\it  compatible NEC and CRF}  for the case of 
        $k=6$ extra dimensions.  For any point $\{w_0, w_a\}$ lying within ${\cal C}$, one can find at least one choice of 
$\{w_b,\zeta_{acc}\}$ 
        that satisfies the NEC and the dark energy theorems and that is matter-dominated in the past.  Conversely,
for points outside ${\cal C}$, at least one of these conditions is violated for every choice of $\{w_b,\zeta_{acc}\}$.}
    \end{center}
    \end{figure}

For $\zeta$, there is only one free parameter that, without loss of generality, can be chosen to be $\zeta_{acc}$, the value when $a=a_{acc}$.  
For a given $\{w_0, w_a, w_b,\zeta_{acc}\}$, the behavior of  $\zeta$ is 
determined by integrating (\ref{e:ZetaDIE}); see Ref.~ \cite{Wesley:2008fg} for details.
We define the set of ``theoretically allowed'' values of $\{w_0, w_a, w_b,\zeta_{acc}\}$ to be those 
that satisfy all of the following conditions:
\begin{itemize}
\item $\wdark \ge -1$ for all $a<2$, so that the dark energy component satisfies the null energy condition for all times;
\item $\wdark \le 0$ as $a \rightarrow 0$, so the universe is sure to be matter- and radiation-dominated 
at very early times;
\item the constraint equations (\ref{e:CRF6ZetaDIE}) and (\ref{e:CRF6ZetaID}) are obeyed, so the theory is compatible with the 
dark energy theorems.
\item $\Omega_{DE}$, the ratio of the dark energy density to the critical density equals 0.74, consistent with current observational constraints for $\wdark$  near $-1$.  
\end{itemize} 
We then project this set of points in the four-dimensional parameter space into the two-dimensional $w_0-w_a$ subspace.  A point $\{w_0, w_a\}$ in this subspace is labeled ``compatible with NEC and CRF'' if there is at least one choice of $\{ w_b,\zeta_{acc}\}$ such that $\{w_0, w_a, w_b,\zeta_{acc}\}$  satisfies all the conditions above. As shorthand, the set of compatible points is labeled ${\cal C}$.  See Figure~\ref{f:NEC_v_SEC}.

Note that the NEC/CRF compatible region in Figure \ref{f:NEC_v_SEC} 
includes the ``de Sitter" point $(w_0,w_a) = (-1,0)$, which appears to 
contradict the claim in  Section~\ref{s:basiceqs} that a universe with constant  $\wtot < w_{\rm trans}$ violates the dark energy theorems.  There is no real 
inconsistency, though: a true 4d de Sitter universe with $\wtot=-1$ (or, equivalently, $w_0=-1, \, w_a=w_b=0$) for all time is incompatible with the dark energy theorems.   However, it is also possible to choose $w_b$ non-zero such that $(w_0,w_a)$ is transiently equal to $(-1,0)$ in the present epoch, and yet have $\wtot$ increase to values greater than $w_{trans}$ in the past and future, in keeping with the dark energy theorems.

We next add the observational limits on the time-variation of $\wdark$ and $G$.  For $\wdark$, the current limits on $w_0$ and $w_a$ and the DETF projections for future experiments are used, as reviewed in Appendix \ref{aa:BB}.  
The current variation of $G$ can be computed from the value of $\zeta$ 
by 
\begin{equation}
\frac{\dot G}{G}\bigg{|}_{\rm today} = ( -7.4 \times 10^{-11} \; \text{yr}^{-1}) \times \zeta_0,                                                                                                                              
\end{equation}
where $\zeta_0$  denotes the present value of $\zeta$; the Hubble parameter is taken to be 
$H = 72$ km/s/Mpc; and we have used (\ref{e:ZetaDotLnG}) to relate the variation in $G$ to
$\zeta$.
The present-day instantaneous constraint on $\dot G/G$ is \begin{equation}\label{e:dotLnG}
\frac{\dot G}{G}  = (0 \pm 5) \times 10^{-12} \; \text{yr}^{-1} 
\end{equation}
which results from various experimental studies described in Appendix \ref{aa:AA}.  Integrating (\ref{e:ZetaDotLnG}) gives a formula that relates $\zeta(N)$ to  the secular variation of $G$ between two times $t_1$ and $t_0$ 
\begin{equation}\label{e:ZetaGratio}
\frac{G(t_1)}{G(t_0)} = \exp  \int_{N(t_1)}^{N(t_0)} \zeta(N) \, \text{d} N
\end{equation}  
For the secular variation of $G$, the current bound is
\begin{equation}\label{e:Gratio}
\frac{G_{BBN}}{G_0} = 1.00^{+\, 0.20}_{-\, 0.16},
\end{equation}
where $G_{BBN}$ and $G_0$ are the values of $G$ at big bang nucleosynthesis (BBN) and at the present, respectively.  Further details about the constraints on secular variation of $G$ may be found in Appendix \ref{aa:AA}.

Using these constraints, we can define a $\chi^2$ measure
of the agreement between observations and models compatible with
the dark energy theorems. The four parameters $w_0$, $w_a$, $w_b$ and $\zeta_{acc}$ suffice
to predict the time variation of  $\wdark$ and $\dot{G}/G$.   
Then, $\chi^2$ as a function of these four variables is 
\begin{equation}
\label{e:X2}
\chi^2 (w_0,w_a , w_b,\zeta_{acc}) = \left( \frac{w_p+1}{\delta w_p} \right)^2 + \left( \frac{w_a}{\delta w_a} \right)^2 + \left( \frac{\dot G/G}{\delta[\dot G/G]} \right)^2 + \left( \frac{G_{BBN}/G_0 -1 }{\delta[G_{BBN}/G_0]} \right)^2,
\end{equation}
where $\delta w_0$, $\delta w_a$, ... are  1$\sigma$ uncertainties,
and $w_p$ is the value of $\wdark$ at the ``pivot'' red shift $z_p=0.3$. We introduce $w_p$ 
because the DETF convention expresses experimental constraints in terms of $w_0$ 
and $w_p$, as described in Appendix \ref{aa:BB}.  In terms
of our model parameters,  $w_p$  is given approximately by
\begin{equation}
w_p = w_0+ 0.231 w_a + 0.0533 w_b
\end{equation}
In our definition of  $\chi^2$, we have assumed that future
 observational data returns a result that best fits $(w_0,w_a) = (-1,0)$ in the 
$w_0$-$w_a$ plane, and null results for instantaneous and secular variation of 
$G$, albeit with some uncertainty in these values.  
We assume the future data differs from current data only in its progressively
tighter error bars.

We have found it useful to reduce the full
four-variable $\chi^2$ function (\ref{e:X2}) to one which depends on only
two variables:  $w_0$ and $w_a$.  We define this two-variable $\chi^2$ function as
the {\it minimum value of the full $\chi^2$ function (\ref{e:X2}) over all $(w_b, \zeta_{acc})$ in the compatibilty region ${\cal C}$, for the given $(w_0,w_a)$}.   Formally,
\begin{equation}\label{e:X2w0wa}
\chi^2 (w_0,w_a) \equiv \underset{(w_b,\zeta_{acc})}{\text{min}} \; \chi^2 (w_0,w_a , w_b,\zeta_{acc}),
\end{equation}
where the minimization is understood to be over all $( w_b,\zeta_{acc})$
such that $(w_0,w_a,w_b,\zeta_{acc}) \in {\cal C}$.
If we were estimating the values of model parameters, it would be more
appropriate to marginalize over $w_b$ and 
$\zeta_{\rm acc}$ than to seek the minimum $\chi^2$.  
However, we want to determine whether any model
compatible with the dark energy theorems can also be consistent with 
future high-precision experiments which give results supporting a
cosmological constant and with no time-variation in $G$.  
Since we are not
interested in the specific value of any of our model parameters, but instead
in the quality of the best fit, $\chi^2$ minimization is more appropriate
than marginalization.
In the following section, we use the methodology described here to compute $\chi^2(w_0,w_a)$ based on current and near-future
observational constraints.


\section{Results}\label{s:results}

In the previous section, 
the dark energy theorems were shown to forbid  equation of state parameters $(w_0, w_a)$ that lie outside region ${\cal C}$ in Fig.~\ref{f:NEC_v_SEC}.  In this section, we consider how observational constraints on $\wdark$ and $\dot G/G$ can further limit and perhaps rule out region ${\cal C}$ itself and, hence, the entire family of NEC/CRF models.  
Current data is consistent with $\wdark =-1$ and $\dot{G}/G =0$; as noted in the previous section, we assume for the purposes of this study that improved measurements will continue to point to the same conclusions about $\wdark$ and $\dot G/G$ in the future, but with smaller uncertainties.    We compute the minimum $\chi^2(w_0,w_a)$ statistic in (\ref{e:X2}) based on this assumption.  A model will be considered ``ruled out'' if  $\chi^2(w_0,w_a)$ exceeds $3 \sigma$  ($99.7\%$ for two parameters).

    \begin{figure}
    \begin{center}
    \includegraphics[width=6in]{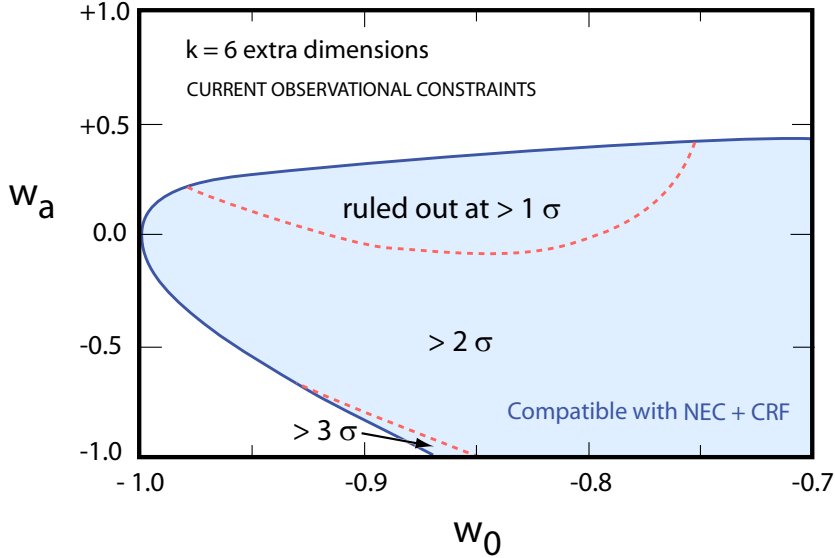}
    \caption{\label{f:nec6_curr}
        Current constraints on the $(w_0,w_a)$ plane for
        NEC/CRF family of models with $k=6$ extra dimensions.
        This plot combines present constraints on both dark energy parameters 
        and $\dot G$, derived using the techniques described in the 
        manuscript.  
        The contours are generated by choosing the model
        which best agrees with experimental constraints 
        amongst all models which
        obey the conditions of the dark energy theorems 
        at each value of $(w_0,w_a)$.    
    }
    \end{center}
    \end{figure}

\begin{figure}
    \begin{center}
    \includegraphics[width=6in]{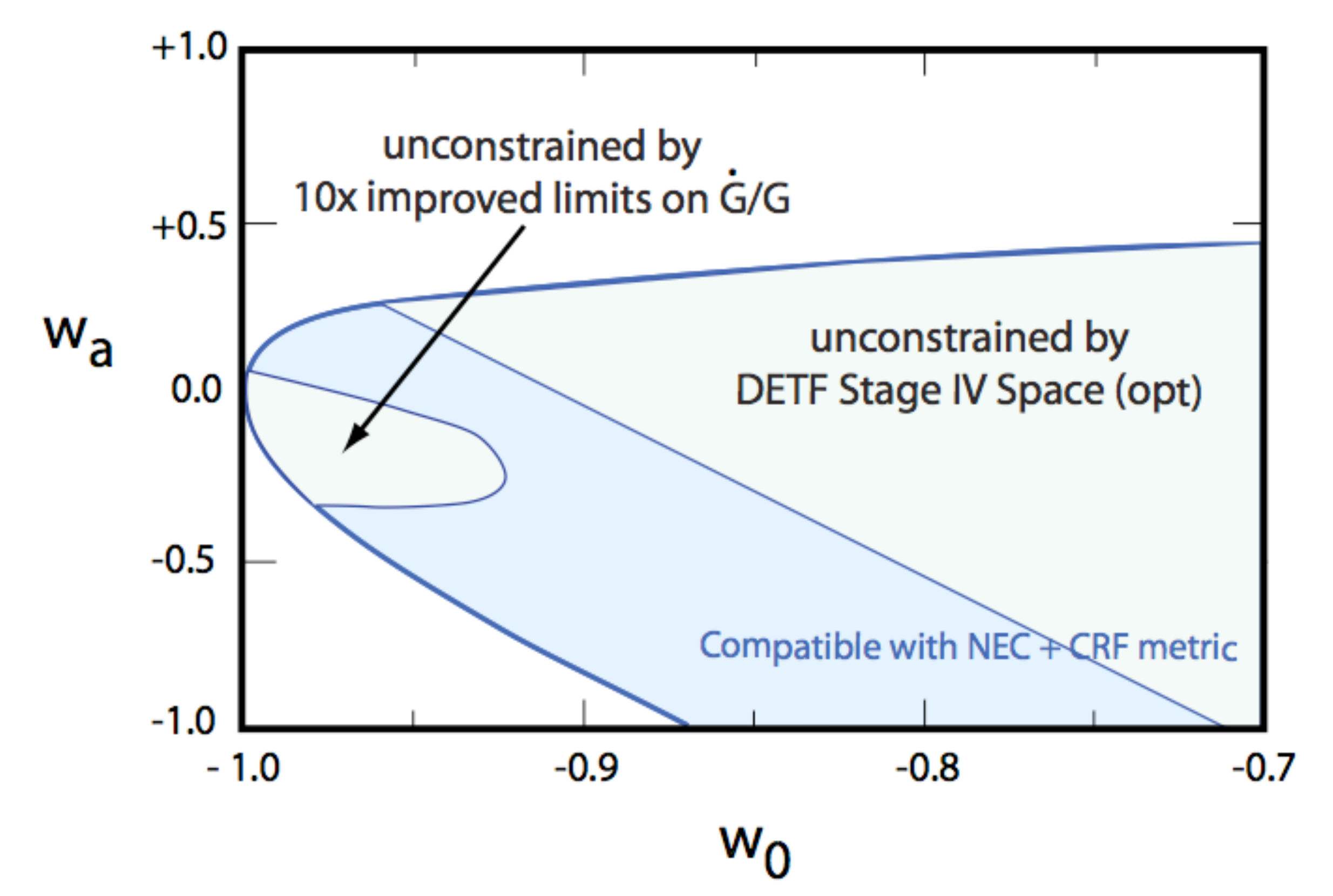}
    \caption{\label{f:nec6_de_or_Gdot}
        Imposing the constraints from dark energy or $\dot G/G$ independently
        does not constrain the NEC/CRF family of models.  Even a very
        ambitious dark energy measurement, or a significant improvement
        in pulsar constraints, would leave a region of parameter space which
        is compatible with the NEC/CRF assumption.  Hence, we cannot rule
        out this family of models by considering these constraints 
        independently.
    }
    \end{center}
    \end{figure}

Figure \ref{f:nec6_curr} illustrates  $\chi^2(w_0,w_a)$ in region $\cal{C}$ based on current observations.  Only a small sliver of ${\cal C}$ is ruled out.   
   Figure 
    \ref{f:nec6_de_or_Gdot} shows that improving measurements of dark energy only or of $\dot G/G$ only is
    not powerful enough
 to rule out the entire NEC/CRF family of models.  For example, the figure shows that a substantial range of ${\cal C}$ near the ``de Sitter" point $(w_0,w_a) = (-1,0)$ 
is still allowed
even if the projected sensitivity of the most ambitious and optimistic DETF space-based proposal is achieved.  Similarly, 
   a tenfold improvement in $\dot G/G$  limits from pulsars, with
    no dark energy information included, leaves unconstrained a substantial range of ${\cal C}$ with $w_0 \gtrsim -0.9$.  The key point, though, is that the poorly constrained regions of ${\cal C}$ for the two  measurements
    do not overlap, suggesting that a combination of the two can be effective in ruling out all of the NEC/CRF family of models.

\begin{figure}
    \begin{center}
    \includegraphics[width=6in]{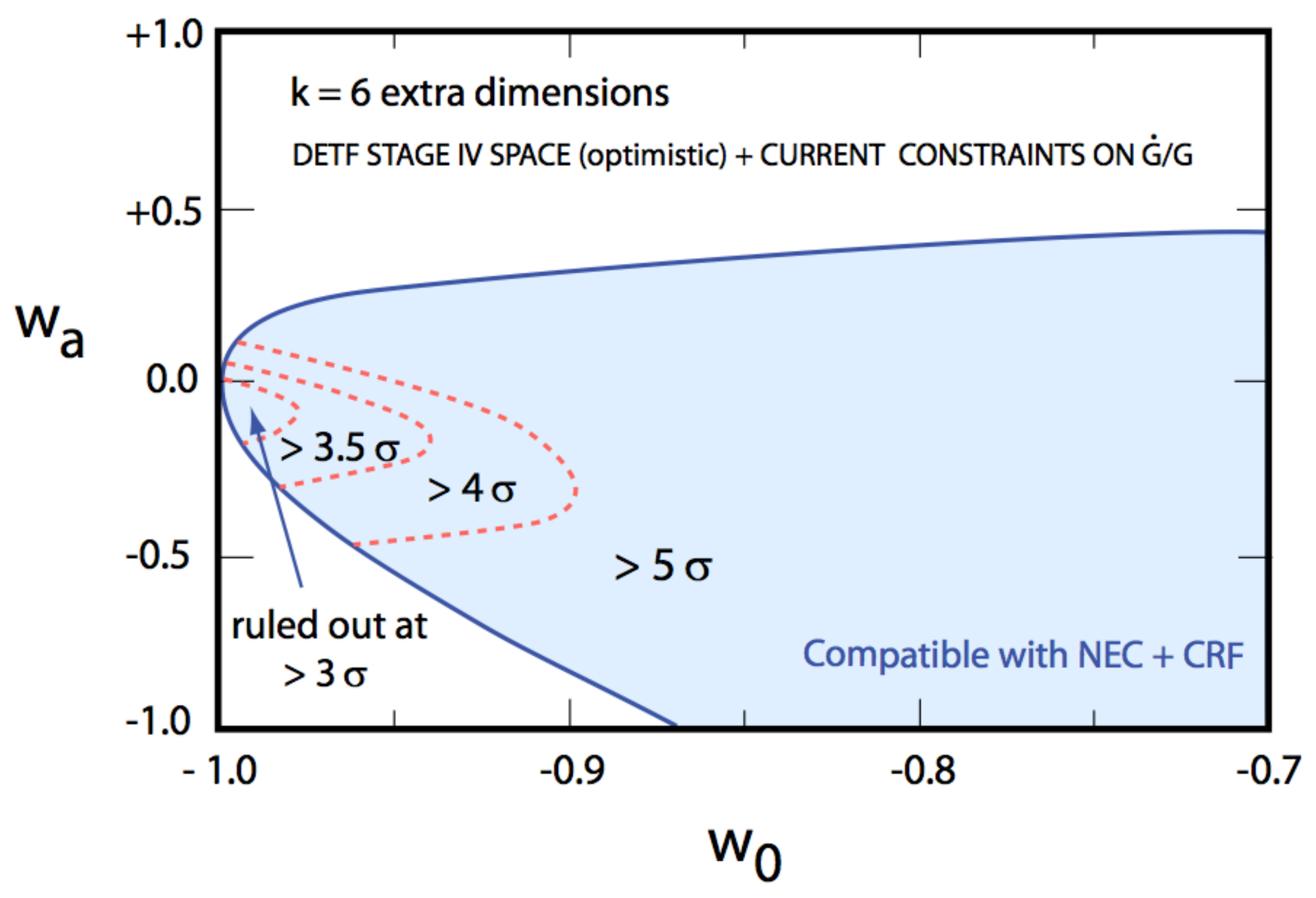}
    \caption{\label{f:nec6_super_de}
        Projected
        constraints after a Stage IV space-based mission (optimistic) for
        models obeying the NEC with CRF-type metrics in $k=6$ extra dimensions.
        With dark energy measurements alone, very ambitious experiments are 
        needed to exclude the full NEC-compatible region, even when
        current constraints on $\dot G/G$ are included.  
        The Stage IV space-based mission is the only DETF 
        scenario which excludes the entire plane to $>3\sigma$ for
        both pessimistic and optimistic projections.
    }
    \end{center}
    \end{figure}

For example, constraining NEC/CRF models by improving dark energy constraints 
only and 
combining them with {\it current}  $\dot G/G$ constraints can rule out the entire NEC/CRF
family of models, at the cost of a very ambitious dark energy mission.
    Figure~\ref{f:nec6_super_de} shows 
    $\chi^2(w_0,w_a)$ assuming the most ambitious and optimistic DETF concept,  a {\it space-based} Stage IV proposal.  All of ${\cal C}$ is ruled out by more than $3 \sigma$.  
    (Stage III and Stage IV 
{\it ground-based} missions may be able to rule out the entire plane if the 
optimistic DETF projections hold true 
(at $\sim 3.1\sigma$ and $\sim 3.4\sigma$, respectively), 
but  they would fall short for DETF pessimistic projections.)   Although the figure is not shown here, improving constraints on $\dot{G}/G$ only and combining with the current limits on $\wdark$ is insufficient to rule out all of ${\cal C}$.  In short, improving only one of the two measurements is a difficult approach, at best,
for ruling out the NEC/CRF family of models.

\begin{figure}
    \begin{center}
    \includegraphics[width=6in]{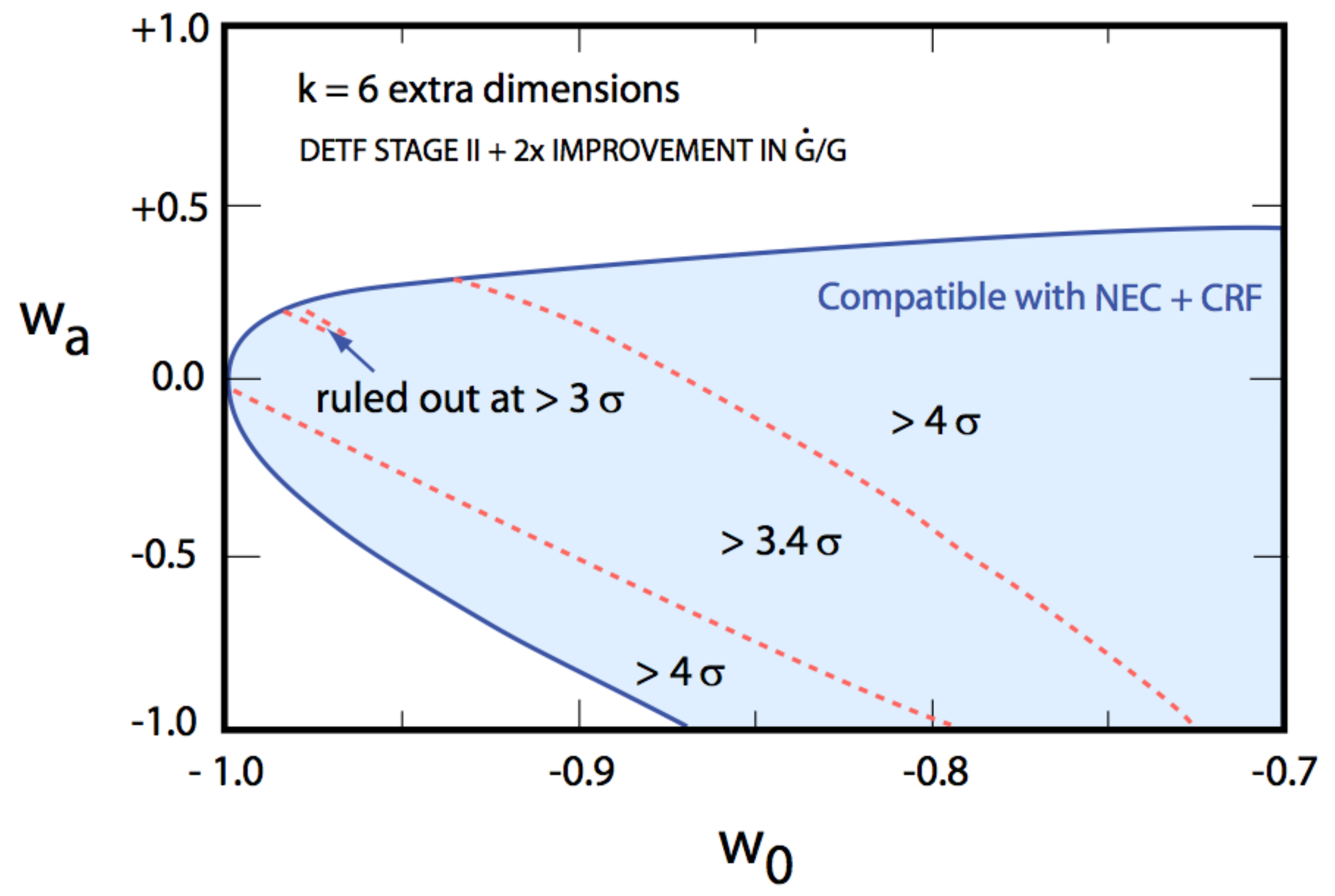}
    \caption{\label{f:nec6_mod_de}
        Projected
        constraints after a DETF Stage II dark energy measurement and
        a twofold improvement in $\dot G/G$ bounds,  for
        models obeying the NEC with CRF-type metrics in $k=6$ extra dimensions.
        This family of models is excluded at the same level as with
        a Stage IV dark energy measurement and current $\dot G/G$ bounds,
        hence modest improvements in constraints on 
        both dark energy and  $\dot G/G$ can 
        do much better than ambitious improvements in dark energy constraints
        alone.
    }
    \end{center}
    \end{figure}

   A less demanding strategy involves modest improvements to measurements of both $\wdark$ and $\dot{G}/G$,  as illustrated in Figure \ref{f:nec6_mod_de}.  This figure 
    shows the exclusion 
    regions with DETF Stage II $\wdark$ measurements, and only a factor of two improvement in the
    current value of $\dot{G}/G$.  The entire range of ${\cal C}$ can be ruled at  $\gtrsim 3.4\sigma$.

\section{\label{s:conclusions}Discussion}

The results of the previous section demonstrate that it is possible to test and possibly rule out an entire class of extra-dimensional models in a way that does not depend on the size of the extra dimensions.   The approach relies on the fact that the expansion of the universe is accelerating today and that acceleration causes the volume of the compactified dimensions to change in cases where the theory satisfies a classical 
energy condition. 
Instead of probing the compactified dimensions directly, the approach is to test for the effect of time-variation of the compactified dimensions on $G$ and $\wdark$.  A key advantage of this approach is that the time variation required by dark energy theorems does not depend on the size of the extra dimensions.  

For the purposes of illustration, we have focused on CRF models that obey the NEC because they are common to many string- and M-theoretic constructions.  We have demonstrated that current observations allow a significant range of these models, but that improvements in the measurement of $\dot G/G$ and $\wdark$ anticipated over the next few years can rule them out entirely.  
To accomplish the task by 
improved dark energy constraints alone
 would require the DETF's most ambitious  
  Stage IV plan combined with current measurements of $\dot G/G$.

\begin{figure}
    \begin{center}
    \includegraphics[width=6in]{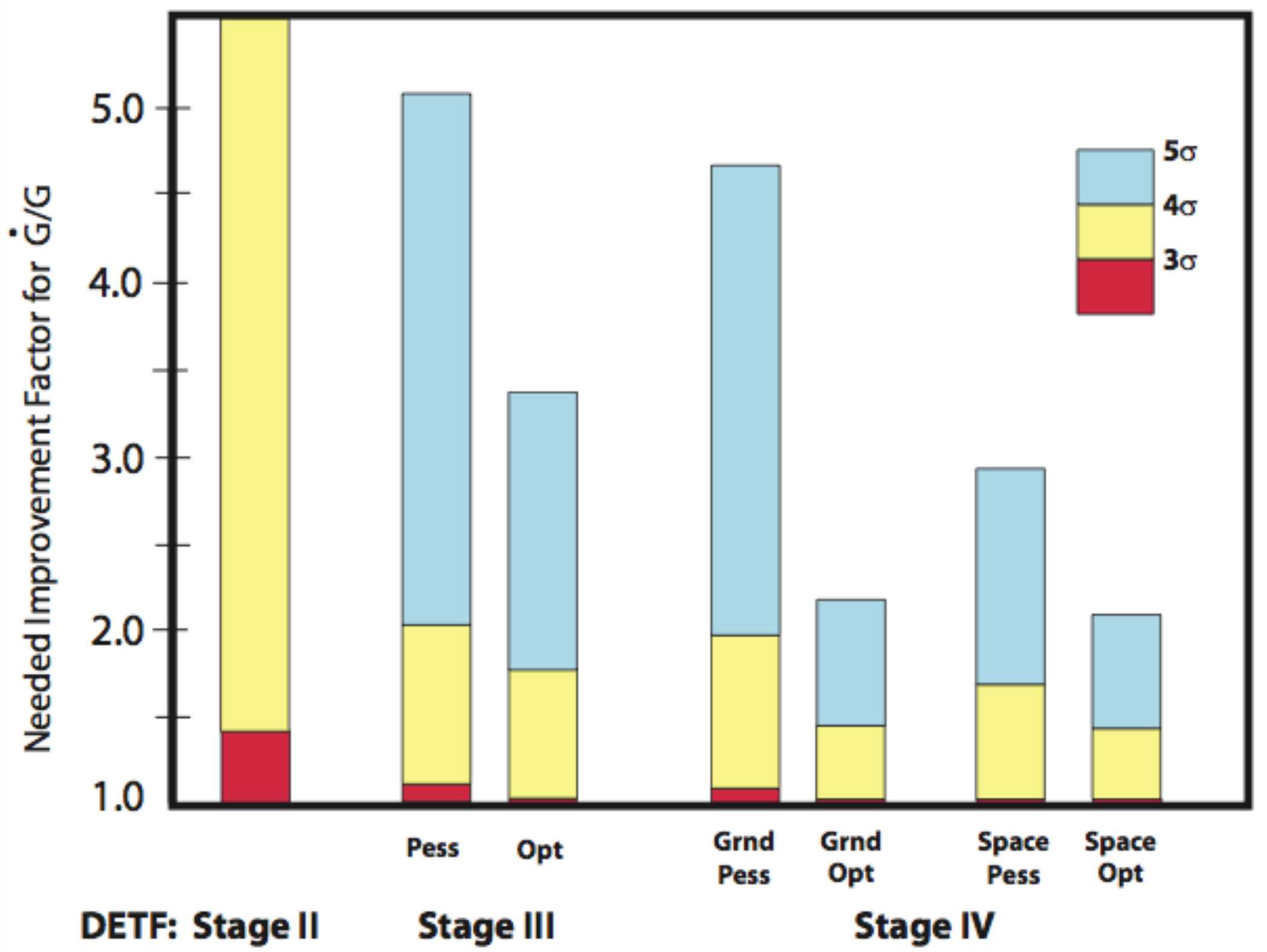}
    \caption{\label{f:requiredPulsar}
        The improvements in $\dot G/G$ constraints required to rule
        out NEC-satisfying models with CRF-type metrics and $k=6$ extra
        dimensions, as a function of the available dark energy constraints.
        For each DETF stage, the bars show the improvement in pulsar constraints
        that would lead to exclusion of this family of models at the
        3$\sigma$, 4$\sigma$, and 5$\sigma$ levels.  By contrast, this family
        cannot be ruled out (even at the 1$\sigma$ level) from dark energy
        information alone, even with constraints from extremely ambitious 
        measurements.
        }
    \end{center}
    \end{figure}

{\it The most cost-effective approach,, though, is to combine a modest improvement in $\dot G/G$ and $\wdark$ limits, which can 
impose constraints tighter than those obtained from a very high-precision dark energy measurement alone.}  Fig.~\ref{f:requiredPulsar} summarizes how different combinations of improved measurements of $\dot{G}/{G}$ from pulsar timing measurements and dark energy experiments described by the DETF can rule out the entire range of $k=6$-dimensional CRF models obeying the NEC at the $3 \sigma$ level or higher.

A similar approach can be applied to test other metrics and/or other energy conditions.  For example, we have carried out a similar analysis
for the family of models in which the extra-dimensional theory satisfies the {\it strong energy condition} (SEC) and is described by curved metrics of the form:
\begin{equation}\label{e:TDmetric}
\text{d} s^2 = e^{2\Omega(t,y)} g^{\rm FRW}_{\mu\nu}(t,x) \, \text{d} x^\mu \text{d} x^\nu + h_{\alpha\beta}(t,y) \, \text{d} y^\alpha \text{d} y^\beta
\end{equation}
where $g^{\rm FRW}_{\mu\nu}$ is a flat Friedmann-Robertson-Walker metric; and   the extra-dimensional metric $h_{\alpha\beta}$ and warp factor $\Omega$ can be 
time-dependent.  Unlike the CRF family, in (\ref{e:TDmetric}) we
allow the extra-dimensional metric $h_{\alpha\beta}$ to have arbitrary Ricci
curvature.
This family of models satisfies a set of dark energy theorems that is different from the theorems for NEC/CRF family.  We find that, as in the case of the NEC/CRF metrics, the SEC/Curved family of models is not ruled out by current constraints on $\wdark$ and $\dot{G}/G$, but DETF Stage II limits combined with current constraints on $\dot{G}/G$ is sufficient to rule out the entire family of models.

We expect to be able to extend our approach to yet more combinations of  metrics and energy conditions.  Even for the NEC/CRF family of models, it may be possible to derive additional dark energy theorems. The current theorems specify conditions that are necessary to have cosmic acceleration and still satisfy the energy conditions, but they are not sufficient.  The current theorems were selected because they are the simplest to prove, as discussed in Ref.~\cite{Steinhardt:2008nk}.  However, there may be stronger theorems that combine with observational constraints of $\wdark$ and $\dot G/G$ to produce much more stringent constraints on extra-dimensional models.       

\appendix

\section{\label{aa:AA} Constraints on time-variation of $G$}

\begin{figure}
    \begin{center}
    \includegraphics[width=6in]{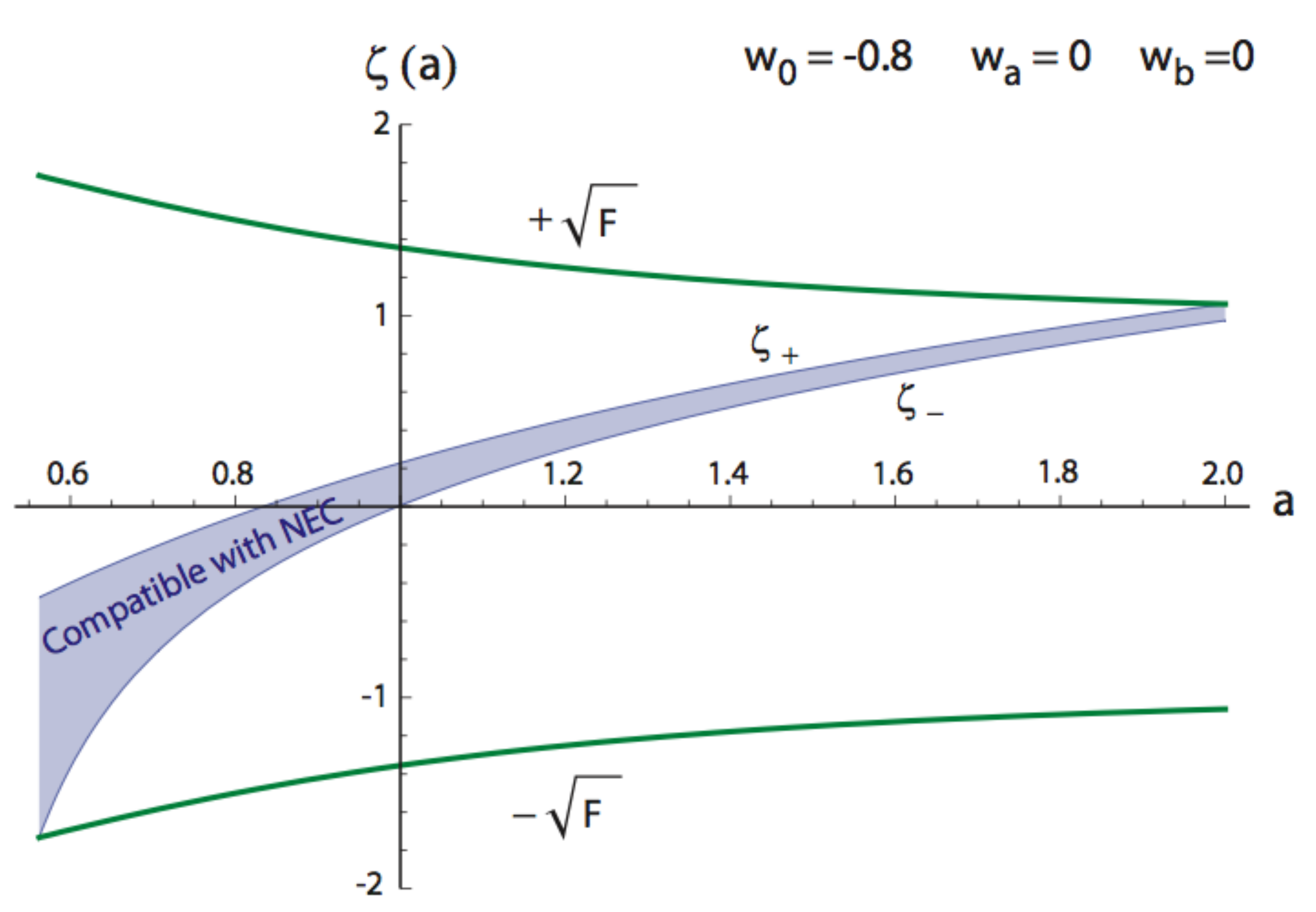}
    \caption{\label{f:zetaDemoPlot} 
        The constraints on $\zeta$ imposed by the dark energy theorem inequalities (\ref{e:ZetaDIE}) and (\ref{e:ZetaID}) for a 
        flat matter-quintessence model with $\Omega_{\rm dark} = 0.74$ and 
        $\wdark = -0.8$ being constant, for which $w_{\rm tot}$ passes through 
        $-1/3$ 
        at $z = 0.78$.  The curves are computed for 
        NEC-satisfying models with CRF-type metrics
        and $k=6$ extra dimensions.  The upper and lower 
        curves are the limits $\pm \sqrt{F}$ obtained from (\ref{e:ZetaID}).  
        The curve $\zeta_-$ is the solution obtained by saturating 
        (\ref{e:ZetaDIE}) and using the most negative initial value for $\zeta$ 
        allowed by (\ref{e:ZetaID}).  The curve $\zeta_+$ is obtained in the 
        same way, but uses the largest possible initial value for $\zeta$ which 
        does not cross over $\sqrt{F}$ by $a=2$.  All $\zeta (a)$ trajectories
        which are compatible with NEC must lie within the central band.
    }
\end{center}
\end{figure}

In Section \ref{s:method} we described how the higher-dimensional Einstein equations and energy conditions are distilled into the expressions (\ref{e:ZetaDIE}) and (\ref{e:ZetaID}) for
$\zeta \equiv – \dot{G}/({G} H) $.
Since (\ref{e:ZetaDIE}) and (\ref{e:ZetaID}) are inequalities, they cannot be used to predict a unique value for $\zeta$, but they can place tight constraint the range of values which $\zeta$ can assume.  This in turn places constraints on the allowed variation of $G$.  The basic technique is illustrated in Figure \ref{f:zetaDemoPlot}.  The plot assumes a cosmological model in which $\wdark = -0.8$ and is constant in time.  The upper and lower black curves in Figure \ref{f:zetaDemoPlot} are $\pm \sqrt{F}$.   If the extra-dimensional model obeys the conditions of the theorems, then $\zeta$ must remain in the region between the two outer curves, according to (\ref{e:ZetaID}).  This immediately constrains the present value of $\zeta$, and hence $\dot G/G$. 
For the NEC/CRF family of models considered in this paper, 
 $F = 9(1+\wtot)/2$ today is of order unity, and so this initial constraint tells us that $G$ should vary by no more than order unity per Hubble time.

To determine the constraints derived from 
the differential inequality (\ref{e:ZetaDIE}), it suffices to consider
trajectories $ \zeta_\pm$ that saturate the inequality, 
as illustrated in Figure \ref{f:zetaDemoPlot}.    Since this differential equation is first-order, the trajectories obtained with different initial values of $\zeta$ never cross each other, so trajectories that satisfy the inequalities are bounded by $\zeta_{\pm}$.  As a practical matter, we impose initial conditions for $\zeta$ at the beginning of the accelerating epoch, when $w=-1/3$, and integrate $a=2$, figuring that our simple parameterization of dark energy should not be trusted for a much wider range of $a$.  

We define $\zeta_-$ as the solution to (\ref{e:ZetaDIE}) which saturates the inequality and uses the smallest allowed initial value of $\zeta$ when $\wtot=-1/3$.  By an "allowed" initial value, we mean one for which the trajectory resulting from integrating (\ref{e:ZetaDIE}) never leaves the envelope $\pm \sqrt{F}$ before the end of the integration interval.  In some cases, such as the one illustrated in Figure \ref{f:zetaDemoPlot}, the initial value for $\zeta_-$ is at the lower edge of the $\pm \sqrt{F}$ envelope, at $\zeta = -\sqrt{F}$.  In other cases, a larger initial value may be necessary to avoid leaving the $\pm \sqrt{F}$ envelope before the end of the integration interval.
The function $\zeta_+$ is defined in a parallel fashion: it is the trajectory defined by (\ref{e:ZetaDIE}) which uses the largest possible initial value of $\zeta$ and does not leave the $\pm \sqrt{F}$ envelope before the end of the integration interval.

Using the two functions $\zeta_\pm$ we can put much tighter constraints on the allowed behavior of $\zeta$.  The solutions $\zeta_\pm$  divide the allowed region between $\pm \sqrt{F}$ into three subregions:

The $-\sqrt{F} < \zeta < \zeta_-$ region is forbidden by the energy condition.  We can show this by contradiction. Suppose there is a curve $\bar{\zeta}(a)$ which reaches a point $(\zeta_1,a_1)$ located in this region but satisfies the conditions of the theorems.  Then at the beginning of acceleration, $\bar{\zeta} > -\sqrt{F}$ by (\ref{e:ZetaID}).  By continuity, the curve $\bar{\zeta}(a)$ must have crossed the curve $\zeta_-$ at some $a_\star < a_1$, and in particular gone from above the curve to below.  This is a contraction, for at the crossing, the slope of $\bar{\zeta}$ is less than that of $\zeta$, and since $\zeta = \bar{\zeta}$ at the crossing point, (\ref{e:ZetaDIE}) could not have been satisfied.  A similar argument applies to 
$\zeta_+ < \zeta < +\sqrt{F}$.  Therefore any function $\bar{\zeta}$ which satisfies both inequalities (\ref{e:ZetaDIE}) and (\ref{e:ZetaID}) is confined to the region $\zeta_- \le \zeta \le \zeta_+$.  Using the relationships (\ref{e:ZetaDotLnG}) and (\ref{e:ZetaGratio}), this constraints on $\zeta$ translate into constraints  on the allowed values of the ratio between $G$ at the beginning of the accelerating epoch and today, and the present value of $\dot G/G$.  

For example,  in the case shown in Figure \ref{f:zetaDemoPlot}, the present values of $\zeta_-$ and $\zeta_+$ today are approximately $5.87 \times 10^{-3}$ and $2.26 \times 10^{-1}$, corresponding to 
\begin{equation}
-1.68 \times 10^{-11} \; \text{yr}^{-1} \le \frac{\dot G}{G}\bigg{|}_{\rm today} \le -4.33 \times 10^{-13} \; \text{yr}^{-1}
\end{equation}
where the upper limit comes from $\zeta_-$ and the lower limit from $\zeta_+$.
This entire range is essentially consistent with the current experimental limits (\ref{e:dotLnG}) at roughly $3\sigma$.  There is a gap between the upper limit and zero, which means that $G = $ constant is not consistent with the NEC/CRF family of  models.  This gap could be explored at $3\sigma$ with a significant improvement of the instantaneous $\dot G/G$ constraints (\ref{e:dotLnG}) by roughly a factor of $\sim 35$.  The limits from secular change in $G$ over the accelerating epoch are
\begin{equation}
0.678 \le \frac{G_{BBN}}{G_0} \le 0.929
\end{equation}
where the lower limit corresponds to $\zeta_+$ and the upper one to $\zeta_-$.

\subsection{Observational constraints on the instantaneous variation of $\dot G / G$}\label{aa:Ginst}

    There is a long history of searches for variation of Newton's constant $G$ 
    (for a recent review, see \cite{Uzan:2002vq}).  In recent years interest in 
    measuring the variation of $G$ (as well as other fundamental constants)  
    has been driven by ideas in higher-dimensional unification and modified 
    theories of gravity which may explain cosmic acceleration.

To be useful in our analysis, it is important that we use constraints
on the time-variation of $G$ that do not make specific assumptions about the functional form of the time-dependence. 
For example, some constraints on the instantaneous variation of $\dot G / G$ are obtained by assuming $G(t) \sim t^\beta$. 
For some special functional forms, such as this, one can obtain very strong observational limits on $\dot{G}/G$ by applying constraints on the value of $G$ in distant past, ({\it e.g.,} from big bang nucleosynthesis).  We cannot use such constraints, because there is no guarantee that the variation of $G$ predicted by the dark energy theorems will follow any particular simple form.  Similarly, we cannot use constraints based on stellar evolution that average over one or more Gyrs to place a limit on the current instantaneous variation of $G$ since the current value of $\dot{G}/G$ may be very different from the average over the last Gyr. 

We also must avoid  constraints which depend on a particular modification of gravity.  By assuming a particular gravitational theory, it is sometimes possible to obtain tight constraints on theory parameters by other measurements, and then use these parameters to derive a constraint on the current value of $\dot G / G$.  One common example is to assume Brans-Dicke theory:  constraining the Brans-Dicke parameter $\omega_{\rm BD}$ by, for example, measuring the Shapiro delay can be used to provide an indirect constraint on the current variation of $G$.  We must discard such constraints because the assumption of a particular gravitational theory is always in the background.

The ideal constraints on the current variation of $G$ are those which use data over very short timescales -- such as a few decades -- that are extremely small in comparison with cosmological timescales, and make minimal assumptions about 
the gravitational theory being tested.  No test can ever be completely model-independent, but assuming Einstein gravity is a conservative assumption which leaves us at little risk for being misled.  

    Solar system constraints are ideal for bounding the variation of $G$, 
    due to the short measurement timescales and nearly Newtonian regime.  
    An early constraint on the variation of $G$ from analyzing ranging data to 
    the Viking Mars lander \cite{Viking} gave a constraint of
    $ \dot G / G  = (2 \pm 4) \times 10^{-12} \; \text{yr}^{-1}$
    The dominant source of uncertainty was the modeling of asteroid effects on 
    the orbit of Mars.  Using the same data set, but a different asteroid model,
     others have obtained slightly weaker limits of 
    $| \dot G / G | < 3 \times 10^{-11} \; \text{yr}^{-1}$ \cite{Viking2} and 
    $| \dot G / G | < 1 \times 10^{-11} \; \text{yr}^{-1}$ \cite{Viking3}, so 
    it is possible that the initial uncertainties may have been underestimated.     
    Measurements using planets in the inner solar system are less sensitive to 
    uncertainties due to the asteroid belt.  An early constraint of 
    $| \dot G / G | < 4 \times 10^{-10} \; \text{yr}^{-1}$ 
    \cite{Shapiro:1971qy} was obtained from radar ranging of Mercury.  This has 
    been sucessively refined  over the intervening decades 
    \cite{ReasShap76,ReasShap78,Shap90}.
    Currently, the constraint 
    \begin{equation}\label{e:SSdotLng}
        \frac{\dot G}{G}  = (0 \pm 2) \times 10^{-12} \; \text{yr}^{-1}
    \end{equation}
    follows from an analysis of a combination of ranging data from Mariner 10, 
    Mercury  and Venus \cite{IPCombo}.

    Pulsar timing measurements have also obtained very stringent bounds on the 
    instantaneous variation of $G$.  In contrast with solar-system tests, 
    post-Newtonian effects in pulsar systems, such as the influence of 
    gravitational binding energy and damping due to the emission of 
    gravitational radiation, cannot be neglected.  There are also theoretical 
    uncertainties involving the composition of the bodies in the pulsar system.   
    Usually in these cases one must take a phenomenological approach to obtain 
    reasonably model-independent bounds.  One approach relates the anomalous 
    pulse period derivative $\delta \dot P$ to the variation of $G$ by 
    $ \delta \dot P / P = - 2 \dot G / G$.   When applied to the Hulse-Taylor 
    pulsar PSR 1913+16, after successive refinements 
    \cite{Damour88,Damour:1990wz,Kaspi:1994hp}
    this gives the bound
    \begin{equation}\label{e:PSRdotLnG}
        \frac{\dot G}{G}  = (4 \pm 5) \times 10^{-12} \; \text{yr}^{-1}
    \end{equation}
    A similar analysis of the pulsar PSR B1855+09 gives the slightly weaker 
    bound $\dot G / G = (-9 \pm 18) \times 10^{-12} \; \text{yr}^{-1}$.  This 
    bound is somewhat more conservative than the one for PSR 1913+16, since the 
    companion is not a neutron star, hence there are fewer composition-dependent   
    uncertainties.

    For our work we adopt the canonical present-day constraint of
    \begin{equation}\label{e:dotLnGcanon}
        \frac{\dot G}{G}  = (0 \pm 5) \times 10^{-12} \; \text{yr}^{-1} \quad 
        \text{(this work)}
    \end{equation}
    This uncertainty envelops both the recent constraints (\ref{e:PSRdotLnG}) 
    from PSR 1913+16  and the inner solar-system tests (\ref{e:SSdotLng}).  This
    is a conservative constraint, for its uncertainty agrees with the pulsar 
    constraint, which is the larger of the two.  Furthermore, the mean value is 
    taken to be zero, in agreement with the solar-system tests and with the 
    simplest theoretical hypothesis that $\dot G / G= 0$ today.

\subsection{Observational constraints on the secular variation of $G$}\label{aa:Gsec}

Another set of constraints on variation of $G$ arise from the integrated variation of $G$ over cosmic history.  Many of the same caveats for constraints on the instantaneous variation of $G$ apply here as well: use constraints which do not assume a specific functional form for the variation of $G$ with time and which do not make strong model-dependent theoretical assumptions.  These constraints bound the ratio between $G_0$, the value of $G$ today, and the value of $G$ at an earlier epoch of cosmic history.

The most precise constraints of the variation of $G$ with cosmic time have come from big bang nucleosynthesis (BBN).   Changing the value of $G$ changes the Hubble expansion rate during BBN, which affects freeze-out temperatures and the rate of various nuclear reactions.  This leads to different predictions for primordial abundances of D, $^3$He, $^4$He, and $^7$Li.  These predictions also depend on the baryon-to-photon ratio $\eta$ and the number of relativistic species during BBN, which is usually parameterized by the number of light neutrino species $N_\nu$.  Allowing $(G,\eta,N_\nu)$ to vary led to the early bound of $0.7 > G_{BBN}/G_0 > 1.4$ \cite{Accetta:1990au}.  Later, using more recent evidence that $N_\nu = 3$ and an independent determination of $\eta$ from cosmic microwave background measurements, this bound was refined to \cite{Copi:2003xd}
\begin{equation}\label{e:GratioBBNX}
\frac{G_{BBN}}{G_0} = 1.01^{+\, 0.20}_{-\, 0.16}
\end{equation}
In this work we take the essentially identical bound
\begin{equation}\label{e:GratioCanon}
\frac{G_{BBN}}{G_0} = 1.00^{+\, 0.20}_{-\, 0.16}\quad \text{(this work)}
\end{equation}
which differs from (\ref{e:GratioBBNX}) by recalibrating the mean value of the ratio to be unity, in accord with the simplest theoretical hypothesis that $G$ has been constant since BBN.

    There is an additional assumption which is specific to our analysis.  As we 
    describe in more detail in Section \ref{s:method}, the dark energy theorems 
    bound the variation of $G$ from the beginning of the accelerating epoch, 
    where $G=G_{acc}$ to the present day, where $G=G_0$.  There are no precise 
    measurements of $G$ at  the transition from deceleration to 
    acceleration, so we will 
    use the ratio of $G_{BBN}/G_0$ as a stand-in for the ratio $G_{BBN}/G_0$.

\section{\label{aa:BB}Constraints on $w_{DE}$}

The discovery that the present universe is accelerating \cite{Riess:1998cb,Perlmutter:1998np} has triggered theoretical efforts to understand this fact, and experimental searches to better characterize the properties of the dark energy that is presumably responsible (for reviews, see \cite{Weinberg:1988cp,Padmanabhan:2002ji,Peebles:2002gy,Copeland:2006wr}).  One simple possibility, consistent with all current data, is that the dark energy is a cosmological constant.  It is also possible that dark energy is dynamical, so that its equation-of-state parameter $\wdark$ varies with time.  Present and future experimental data can constrain this possibility.  A significant problem with applying these constraints is that they usually depend on a specific choice for parameterizing the variation of $\wdark$ with time.

An issue in combining observational constraints and constraints from the dark energy theorems is how to parameterize the time-variation of $\wdark$.       Some studies of supernova data \cite{DiPietro:2002cz,Riess:2004nr} choose a  
    parameterization, first suggested in \cite{Cooray:1999da}, of the form
    \begin{equation}
        \wdark = w_0 + w' z
    \end{equation}
    where $w_0$ is the dark energy equation-of-state parameter 
    today, and $w' = \text{d} w / \text{d} z |_{z=0}$.  With this 
    parameterization, the constraints $w_0 = -1.31^{+\,0.22}_{-\,0.28}$ and 
    $w' = 1.48^{+\,0.81}_{-\,0.90}$ were obtained \cite{Riess:2004nr}, with the 
    errors in each parameter strongly correlated.  If $\wdark$ is assumed to 
    be constant, then the constraints $\wdark = -1.02^{+\,0.13}_{-\,0.19}$ 
    has been obtained \cite{Riess:2004nr}.  Another choice of parameterization, 
    proposed in \cite{Chevallier:2000qy,Linder:2002et}, is
    \begin{equation}\label{e:wDarkLinA}
        \wdark = w_0 + w_a (1-a) = w_0 + w_a \frac{z}{1+z}
    \end{equation}
    which assumes dark energy evolution linear in the scale factor $a$.  This 
    has been used in some recent supernova analyses 
    \cite{Riess:2006fw,Kowalski:2008ez}, along with data from cosmic microwave 
    background (CMB) and baryon acoustic oscillations (BAO). In Ref. 
    \cite{Kowalski:2008ez} this led to correlated errors on $w_0$ and $w_a$ of 
    roughly $\pm 0.2$ and $\pm 0.7$, respectively.  In the same reference, by a 
    flat universe with $\wdark$ constant gives constraints of 
    $\wdark = -1.001 \pm 0.071 \text{(stat)} \pm 0.081\text{(sys)}$.

Analysis of the seven-year Wilkinson Microwave Anisotropy Probe (WMAP) data
has also given constraints on dynamical dark energy \cite{Komatsu:2010fb}.  If
the equation-of-state $\wdark$ is constant, then CMB data alone constrains
$\wdark = -1.10 \pm 0.14$ at $1\sigma$.  Allowing the $\wdark$ to vary 
linearly with $a$ gives the constraints
$w_0 = -0.93 \pm 0.13$ and $w_a = -0.41^{+0.72}_{-0.71}$ at $1\sigma$.
Similar results are obtained from the WMAP five-year analysis
\cite{Komatsu:2008hk}.

    In this work, we draw from these various analyses, and take today's value 
    $w_0$ of the equation-of-state to be
    \begin{equation}
        w_0 = -1.00 \pm 0.13 \quad \text{(this work)}
    \end{equation}
    The uncertainty is the same as the WMAP seven-year analysis, which 
    compares favorably with the the uncertainties in the supernovae analyses.  
    While the WMAP analysis prefers a central value of which is roughly 
    $1\sigma$ from $w_0 = 1$, we have chosen to fix the central value 
    at $-1$ for the purposes of our analysis, based on the notion that 
a cosmological 
    constant is consistent with all data sets and is the simplest assumption.  For the current constraints only, we have chosen to place no 
    limits on $w_a$, since the variety of paremeterizations used in the 
    literature make it difficult to compare this parameter between analyses.  

For projections of future dark energy parameter uncertainties, the Dark Energy Task Force (DETF) report \cite{Albrecht:2006um} is used.  This report introduces the simple linear parameterization in (\ref{e:wDarkLinA}).  However, projected uncertainties are given in terms of two different parameters $(w_p,w_a)$.  The parameter $w_p$ is the value of $\wdark$ at the ``pivot redshift" $z_p \approx 0.3$.  The pivot redshift is the redshift at which a specific experiment can determine the value of $\wdark(z_p)$ with minimum uncertainty.  

The DETF report defines the potential progression in our observational knowledge in terms of four stages:
\begin{itemize}

\item \emph{Stage I}.  Current experiments.

\item \emph{Stage II}. Ongoing experiments related to dark energy.

\item \emph{Stage III}. Near-term, currently proposed projects.  Ref. \cite{Albrecht:2006um} considers ground-based surveys for BAO, cluster lensing, supernovae, and weak lensing.

\item \emph{Stage IV}. Ambitious long-term projects, such as the Large Survey Telescope (LST), Square Kilometer Array (SKA), or Joint Dark Energy Mission (JDEM) 

\end{itemize}
Projections for both pessimistic and optimistic error ellipses in the $(w_p,w_a)$ plane are given for Stages II-IV.  Furthermore, Stage IV projections are given for both ground-only and space-based Stage IV campaigns.  

\begin{table}
\caption{\label{t:DEdata} Data scenarios for dark energy measurements.  These
are taken from the DETF report.  We show the scenario name (used in our
computer code), the uncertainty $\Delta w_p$ in $\wdark$ a the pivot
redshift, and the uncertainty $\Delta w_a$ in $w_a$ at the pivot redshift. }
\begin{ruledtabular}
\begin{tabular}{llll}
Scenario & $\Delta w_p$ & $\Delta w_a$ & Notes \\
\hline
\texttt{node}      &  &  & No dark energy constraints. \\
\texttt{currde}     & $ 0.13$   &   & Current uncertainty, central $w_0=-1$ \\
\texttt{detfII}    & $ 0.045$ & $ 0.66$  & DETF Stage II \\
\texttt{detfIIIp}  & $ 0.031$ & $ 0.36$  & DETF Stage III, pessimistic\\
\texttt{detfIIIo}  &  $ 0.025$ & $ 0.23$  & DETF Stage III, optimistic \\
\texttt{detfIVGp}  &  $ 0.030$ & $ 0.31$  & DETF Stage IV ground, pessimistic \\
\texttt{detfIVGo}  &  $ 0.016$ & $ 0.11$  & DETF Stage IV ground, optimistic \\
\texttt{detfIVSp}  &  $ 0.023$ & $ 0.15$  & DETF Stage IV space, pessimistic \\
\texttt{detfIVSo}  &  $ 0.015$ & $ 0.12$  & DETF Stage IV space, optimistic
\end{tabular}
\end{ruledtabular}
\end{table}

A summary of the various data scenarios we consider is given in Table \ref{t:DEdata}.




\end{document}